\documentclass[superscriptaddress,twocolumn]{revtex4}
\usepackage{amsmath,lscape,epsfig}

\def\ii{\'{\i}}
\def\beq{\begin{equation}}
\def\eeq{\end{equation}}
\def\beqa{\begin{eqnarray}}
\def\eeqa{\end{eqnarray}}
\def\ban{\begin{eqnarray*}}
\def\ean{\end{eqnarray*}}
\def\bi{\begin{itemize}}
\def\ei{\end{itemize}}

\def\d{\mbox{d}}

\begin{document}

\title{Density dependent hadronic models and the relation between neutron
  stars and neutron skin thickness}

\author{S.S. Avancini}
\affiliation{Depto de F\'{\i}sica - CFM - Universidade Federal de Santa
Catarina  Florian\'opolis - SC - CP. 476 - CEP 88.040 - 900 - Brazil}
\author{J.R. Marinelli}
\affiliation{Depto de F\'{\i}sica - CFM - Universidade Federal de Santa
Catarina  Florian\'opolis - SC - CP. 476 - CEP 88.040 - 900 - Brazil}
\author{D.P. Menezes}
\affiliation{Depto de F\'{\i}sica - CFM - Universidade Federal de Santa
Catarina  Florian\'opolis - SC - CP. 476 - CEP 88.040 - 900 - Brazil}
\author{M.M.W. Moraes}
\affiliation{Depto de F\'{\i}sica - CFM - Universidade Federal de Santa
Catarina  Florian\'opolis - SC - CP. 476 - CEP 88.040 - 900 - Brazil}
\author{C. Provid\^encia}
\affiliation{Centro de F\ii sica Te\'orica - Dep. de F\ii sica -
Universidade de Coimbra - P-3004 - 516 - Coimbra - Portugal}

\begin{abstract}
In the present work we investigate the main differences in the
lead neutron skin thickness, binding energy, surface energy and density
profiles  obtained with two different density dependent hadron models.
Our results are calculated within the Thomas-Fermi approximation with two
different numerical prescriptions and compared with results obtained with a
common parametrization of the non-linear Walecka model.
The neutron skin thickness is a reflex of the equation of state properties.
Hence, a direct correlation between the neutron 
skin thickness and the slope of the symmetry energy is found.
We show that within the present approximations the asymmetry parameter for 
low momentum transfer polarized electron scattering is not sensitive to the  
model differences.
\end{abstract}

\maketitle

\vspace{0.50cm}
PACS number(s): {21.65.+f,24.10.Jv,95.30.Tg,26.60.+c}
\vspace{0.50cm}

\section{Introduction}

The relation between neutron star properties which are obtained from adequate
equations of state (EoS) and the neutron skin thickness has long been a topic
of investigation in the literature. The details of this relation and the
important quantities to be discussed have been well established in \cite{tb01},
where it was shown that the difference between the neutron and the proton
radii, the neutron skin thickness, is linearly correlated with
the pressure of neutron matter at sub-nuclear densities. This is so because the
properties of neutron stars are
obtained from appropriate EoS whose symmetry energy depends on the density and
also controls the size of the neutron skin thickness in heavy and
asymmetric nuclei, as $^{208}$ Pb, for instance. It is important to remember
that the EoS in neutron stars is also very isospin asymmetric due
to the  $\beta$- equilibrium constraint.

Hence, isospin asymmetry plays a major role in the
understanding of the density dependence of the symmetry energy and the
consequences it may arise \cite{steiner}.
In \cite{hp01,piek06} it was shown that the models
that yield smaller neutron skins in heavy nuclei tend to yield smaller neutron
star radii due to a softer EoS.

Neutron stars are believed to have a solid crust formed by nonuniform neutron
rich matter in $\beta$-equilibrium above a liquid mantle. In the inner crust
nuclei coexist with a gas of neutrons which have dripped out. The properties
of this crust as, for instance, its thickness and pressure at the crust-core
interface depend a lot on the density dependence of the EoS used to describe
it \cite{haen00,piek06}. On the other hand, it is well known \cite{chom04,
inst04} that the existence of
phase transitions from liquid to gas phases in asymmetric nuclear matter (ANM)
is intrinsically related with the instability regions which are limited by
the spinodals. Instabilities in ANM described within relativistic mean field
hadron models, both with constant and density dependent couplings at zero and
finite temperatures have already been investigated \cite{inst04}
and it was shown that the main differences occur at
finite temperature and large isospin asymmetry close to the boundary of the
instability regions. In  neutral neutron-proton-electron (npe) matter the
electrons are also included.
In a thermodynamical calculation the instabilities almost completely
disappear due to the  high electron Fermi energy \cite{inst062}.

 However,  in a  dynamical calculation which includes the Coulomb
  interaction  and allows for independent neutron, proton and electron
fluctuations \cite{inst06,coletivos}, it is seen that  the electron
dynamics tends to restore the short wavelength instabilities although moderated
by the high electron Fermi energy.

Moreover, it is also known that the
liquid-gas phase transition in ANM can lead to an isospin distillation
phenomenon,  characterized by a larger proton fraction in the liquid
phase than in the gas phase. This is due to the repulsive isovector
channel of the nuclear interaction \cite{xu00,ducoin06,chomaz}.

In a recent work  the spinodal section and
related quantities, as the neutron to proton density fluctuations responsible
for the distillation effect,  has been studied within
different relativistic models \cite{inst062}.
 It was shown
that the  distillation effect within density dependent relativistic
models decreases
with  density  above a nuclear density of $\sim 0.02-0.03$ fm$^{-3}$,
a result similar to the one obtained with the SLy230a parametrization
of   Skyrme interaction \cite{chabanat} and contrary to the results found
with the more common relativistic parametrizations with no density
dependent coupling parameters.   In the last case the distillation
  effect becomes always larger as the density increases.

Also, the behavior of the symmetry energy obtained with density
dependent models is closer to what one obtains with non-relativistic models
than with other relativistic models with constant couplings \cite{inst04}.
In an attempt to understand this behavior, a comparison between
the non-relativistic Skyrme effective force and relativistic mean field models
at subsaturation densities was performed \cite{comp}.  It was shown
  that the relativistic models could also be reduced to an energy
  density functional similar to the one describing the Skyrme interaction.

There have already been some efforts in order to compare
nuclear matter and finite nuclei properties obtained both with relativistic
and non-relativistic models \cite{bao-li,ring97} but there is no clear
or obvious explanations for the differences.
At very low densities both, the relativistic and the
non-relativistic approaches predict a non-homogeneous phase
commonly named {\em pasta phase}, formed by a competition between the
long-range Coulomb repulsion and the short-range nuclear attraction
\cite{pasta}.

Based on the above arguments, it is very important that an
accurate experimental measurement of the neutron skin thickness is
achieved.  This depends on a precise measurement of both
the charge and the neutron radius. The charge radius is already known
within a precision of one  percent for most stable nuclei, using the
well-known single-arm and non-polarized elastic electron
scattering technique  as well as the spectroscopy of muonic atoms
\cite{vries} .
For the neutron radius, our present knowledge has
an uncertainty of about 0.2 fm \cite{horo}. However, using
polarized electron beams it is possible to obtain the neutron
distribution in nuclei in a fairly model independent way, as first
discussed in \cite{Don} and, as a consequence, to obtain the
desired neutron radius.
In fact, the Parity Radius Experiment
(PREX) at the Jefferson Laboratory \cite{prex} is currently
running to measure the $^{208}$Pb neutron radius with an accuracy
of less than 0.05 fm, using polarized electron scattering.

In the present work, we use two different hadronic models that incorporate
density dependence in different ways. The first one, to which we refer next as
the TW model is a density dependent hadronic model with the
meson-to-nucleon couplings explicitly dependent of the density
\cite{original,tw}. In the following it is used to calculate the neutron skin
thickness of $^{208}$Pb, which is a
neutron-rich heavy nucleus. This model was chosen because it is based on a
microscopic calculation, fits well many nuclei properties and, as stated
above, has shown to provide results which are different from the usual
NL3 \cite{nl3} and TM1 \cite{tm1} parametrizations for the non-linear Walecka
model (NLWM),  having a richer density dependence of the symmetry
energy than most of the relativistic nuclear models.
The original motivation for the development of this density dependent hadronic
model \cite{flw,lf} was to reproduce results obtained with the relativistic
Dirac-Brueckner Hartree-Fock (DBHF) theory \cite{DB}. Later the DBHF
calculations for nuclear matter were taken only
as a guide for a suitable parametrization of the density dependence of the
meson-nucleon coupling operators \cite{tw,ring1}.
Moreover, density dependent hadronic models can also be a
useful tool in obtaining EoS for neutron stars even if hyperons are to be
considered \cite{ddpeos}, which is not the case if NL3 or TM1 are used.
Both, NL3 and TM1, can only be used if the EoS is restricted to accommodate
neutrons, protons and the leptons necessary to enforce
$\beta$-stability. Once hyperons are included, the nucleons acquire a negative
effective mass  above $\sim 3-4 \rho_0$ densities
\cite{compact,alex},  where $\rho_0$ is the nuclear saturation density.

The second model, that we refer to as NL$\omega\rho$ model, 
includes non-linear $\sigma-\rho$ and $\omega-\rho$
couplings \cite{hp01,hp2001,bunta,bunta2} which allow to change the density
dependence of
the symmetry energy  of the most common parametrizations of the NLWM that
show  essentially a  linear behavior of the symmetry energy with
  density.  However, the symmetry energy determines the behavior
  of isospin asymmetric matter and therefore is intrinsically related to the
characteristics of the EoS that can describe neutron stars.
Within this model the authors of \cite{hp01} have  shown that
  the neutron skin thickness of $^{208}Pb$ was sensitive to the isovector
channel of the nuclear interaction and there was a correlation between neutron
  skin thickness of nuclei and properties of neutron stars.

For the sake of completeness, the results of the present work, whenever
possible are compared with the results obtained with the NL3 parametrization
of the NLWM, known to describe finite nuclei properties well.

We perform two different numerical calculations to obtain the
$^{208}$Pb properties: a Thomas-Fermi approximation based on the liquid-gas
phase transition developed in \cite{gotas} and a Thomas-Fermi approximation
based on a method proposed in \cite{ring}, where a harmonic oscillator basis
is used. We restrict ourselves to the Thomas-Fermi approximation because, as
we show in the Results section at the end of the paper, for the purpose of
obtaining correct surface energy and neutron-skin thickness, it is almost as
good as the solution of the Dirac equation.

At this point it is worth mentioning that the scalar-isovector $\delta$
mesons, which play an important role in the isospin channel, could also be
incorporated in our work as done in \cite{gaitanos, inst04,inst06} but in
order to make the comparisons among different approximations as simple as
possible, they will be included in a future work. Finally, as we are
interested in nuclei ground state properties, all calculations are performed
at zero temperature.

\section{The TW density dependent hadronic model}

Next we describe the main quantities of the TW model, which has
density dependent coupling parameters. The Lagrangian density reads:
$$
{\cal L}=\bar \psi\left[\gamma_\mu\left(i\partial^{\mu}-\Gamma_v V^{\mu}-
\frac{\Gamma_{\rho}}{2}  \boldsymbol{\tau} \cdot \mathbf {b}^\mu \right. \right.
$$
$$ \left. \left. -e  \frac{(1+\tau_{i3})}{2} A^\mu \right)
-(M-\Gamma_s \phi)\right]\psi
$$
$$
+\frac{1}{2}(\partial_{\mu}\phi\partial^{\mu}\phi
-m_s^2 \phi^2)
-\frac{1}{4}\Omega_{\mu\nu}\Omega^{\mu\nu}$$
\begin{equation}
+\frac{1}{2}m_v^2 V_{\mu}V^{\mu}
-\frac{1}{4}\mathbf B_{\mu\nu}\cdot\mathbf B^{\mu\nu}+\frac{1}{2}
m_\rho^2 \mathbf b_{\mu}\cdot \mathbf b^{\mu}
-\frac{1}{4}F_{\mu\nu}F^{\mu\nu}
\label{lagtw}
\end{equation}
where $\phi$, $V^\mu$, $\mathbf {b}^\mu$ and $A^{\mu}$ are the
scalar-isoscalar, vector-isoscalar and vector-isovector meson
fields and the photon field respectively,
$\Omega_{\mu\nu}=\partial_{\mu}V_{\nu}-\partial_{\nu}V_{\mu}$ ,
$\mathbf B_{\mu\nu}=\partial_{\mu}\mathbf b_{\nu}-\partial_{\nu} \mathbf b_{\mu}
- \Gamma_\rho (\mathbf b_\mu \times \mathbf b_\nu)$,
$F_{\mu\nu}=\partial_{\mu}A_{\nu}-\partial_{\nu}A_{\mu}$
and $\tau_{p3}=1$, and $\tau_{n3}=-1$.
The  parameters of the model are:
the nucleon mass $M=939$ MeV, the masses of
the mesons $m_s$, $m_v$, $m_\rho$, the electromagnetic coupling constant
$e=\sqrt{4\pi/137}$
and the density dependent coupling constants $\Gamma_{s}$,
$\Gamma_v$ and $\Gamma_{\rho}$, which are adjusted in order to reproduce
some of the nuclear matter bulk properties shown in Table \ref{bulk},
using the following parametrization:
\begin{equation}
\Gamma _{i}(\rho )=\Gamma _{i}(\rho _{sat})h_{i}(x),\quad x=\rho /\rho _{sat},
\label{paratw1}
\end{equation}
with
\begin{equation}
h_{i}(x)=a_{i}\frac{1+b_{i}(x+d_{i})^{2}}{1+c_{i}(x+d_{i})^{2}},\quad i=s,v
\end{equation}
and
\begin{equation}
h_{\rho }(x)=\exp [-a_{\rho }(x-1)],  \label{paratw2}
\end{equation}
with the values of the parameters $m_{i}$,
$\Gamma _{i}(\rho_{sat})$, $a_{i}$, $b_{i}$, $c_{i}$ and $d_{i}$,
$i=s,v,\rho $ given in \cite{tw}.
This model does not include self-interaction terms for the meson
fields (i.e. $\kappa =0$, $\lambda =0$ and $\xi=0$ ) as in NL3 or TM1
parametrizations for the NLWM.

The field equations of motion follow from the
Euler-Lagrange equations. When they are obtained, some care has to be taken
since the coupling operators depend on the baryon fields $\bar
\psi$ and $\psi$ through the density.
When the partial derivatives
of $\cal{L}$ are performed relatively to the fields $\bar \psi$
and $\psi$, they yield extra terms due to the functional
dependence of the coupling operators. The new terms are absent in the
usual Quantum Hadrodynamic (QHD, NLWM) models \cite{sw, nl3,tm1}.
The equations of motion for the fields read:

\begin{eqnarray}
(\partial_\mu\partial^{\mu} + m_{\phi}^2)\phi &=&
 \Gamma_s \bar \psi \psi , \label{PHI} \\
\partial_{\nu} \Omega^{\mu\nu} + m_v^2 V^{\mu} &=&
 \Gamma_v \bar \psi \gamma^{\mu} \psi, \label{OME}\\
\partial_{\nu} {\mathbf {B}}^{~\mu\nu} + m_\rho^2
{\mathbf b}^\mu &=&
\frac{\Gamma_\rho} {2} \bar \psi \boldsymbol {\tau} \gamma^{\mu}
\psi, \label{RHO}\\
\partial_{\nu} F^{\mu\nu} &=& \frac{e}{2}\bar \psi
(1+\tau_3 )\gamma^{\mu} \psi, \label{EM}\\
\left[ \gamma_{\mu}(i\partial^\mu -\Sigma^{\mu}) -M^{\ast}
\right] \psi&=&0 ~, \label{DIRAC}
\end{eqnarray}
where $M^{\ast}=M-\Gamma_s \phi$. Notice that in the
equation of motion for the baryon field $\psi$ the vector
self-energy consists of two terms, $\Sigma_{\mu}$ =
$\Sigma^{(0)}_\mu$ + $\Sigma^{R}_{\mu}$, where:
\begin{equation}
 \Sigma^{(0)}_\mu = \Gamma_{\omega} V_\mu
+\frac{\Gamma_{\rho}}{2} \boldsymbol {\tau}\cdot {\mathbf b}_{\mu}
+ \frac{e}{2} (1+\tau_3 ) A_{\mu},
\end{equation}
\begin{equation}
\Sigma^{R}_{\mu}=\left( \frac{\partial \Gamma_v}{\partial
{\rho}} V^\nu j_\nu  + \frac{1}{2}
\frac{\partial \Gamma_\rho}{\partial \rho}  {\mathbf b}_\nu  \cdot \mathbf j_3^\nu  - \frac{\partial
\Gamma_\phi}{\partial {\rho}} \bar \psi \psi \phi\right)
u_\mu ~, \label{REAR}
\end{equation}
where $\Sigma^{(0)}_\mu$ is the usual
vector self-energy, $\hat\rho u_\mu=j_\mu$ with $u^2=1$
$j_\nu=\bar \psi \gamma_\nu \psi$,
$\mathbf j_3^\nu= \bar \psi {\boldsymbol \tau} \gamma^\nu \psi$
and, as a
result of the derivative of the Lagrangian with respect to
$\rho$ a new term appears, $\Sigma^{R}_{\mu}$, which  is
called rearrangement self-energy and has been shown to play an
essential r\^ole in the applications of the theory. This term guarantees the
thermodynamical consistency and the energy-momentum conservation.
For more detailed calculations, at zero and finite temperatures, please refer
to \cite{previous}.

In the static case there are no currents in the nucleus and the spatial
vector components are zero. Therefore, the mesonic equations of motion become:
\begin{equation}
\nabla^2 \phi = m_s^2\phi- \Gamma_s \rho_s,
\label{elphi}
\end{equation}
\begin{equation}
\nabla^2 V_0 = m_v^2 V_0 - \Gamma_v \rho,
\label{elV0}
\end{equation}
\begin{equation}
\nabla^2 b_0 =m_\rho^2 b_0 -\frac{\Gamma_\rho}{2} \rho_3,
\label{elb0}
\end{equation}
\begin{equation}
\nabla^2 A_0 =-e \rho_p,\label{elA0}
\end{equation}
where
$\rho_s=<\bar \psi \psi>$ is the scalar density,
$\rho=\rho_p+\rho_n$, $\rho_3=\rho_p-\rho_n$
and $\rho_p$ and $\rho_n$ are the proton and neutron densities.

\subsection{Thomas-Fermi approximation}

We first define the functional
\begin{equation}
\Omega= E - \mu_p B_p - \mu_n B_n ,
\label{Omega}
\end{equation}
where $E$ is the energy, $\mu_p$ ($\mu_n$) is the
proton (neutron) chemical potential and $B_p$ ($B_n$) is the proton (neutron)
number.
Within the semi-classical Thomas-Fermi approximation,
the energy of the nuclear system with particles described
by the one-body phase-space distribution function
$ f({\mathbf r},{\mathbf p},t)$
at position $\mathbf r$, instant $t$ with momentum $\mathbf p$
is given by
$$E= \sum_i \gamma \int \d^3r  \frac{\d^3p}{(2\pi)^3}\,
f_i({\mathbf r},{\mathbf p},t)
\left(\sqrt{{\mathbf p}^2 + {M^*}^2}
+ {\cal V}_{i}\right)$$
$$ + \frac{1}{2} \int \d^3r  \left[
(\nabla \phi)^2 + m_s^2 \phi^2 -(\nabla V_0)^2 - m_v^2 V_0^2 \right.
$$
\begin{equation}
\left. - (\nabla b_0)^2 - m_\rho^2 b_0^2  - (\nabla A_0)^2 \right]
\end{equation}
where
$$
{\cal V}_{p}= \Gamma_v V_0  + \frac{\Gamma_\rho}{2} b_0 + e A_0\;,
\quad {\cal V}_{n}= \Gamma_v V_0  - \frac{\Gamma_\rho}{2}  b_0 \;,
$$

$\gamma=2$ refers to the spin multiplicity and the
distribution functions for protons and neutrons are
$$f_i=\theta(k_{Fi}^2(r)-p^2), ~~~~~~~i=p,n~.$$
In this approach, the scalar, proton and neutron densities become:
$$
\rho_s(r)= \frac{\gamma}{2\pi^2} \sum_{i=p,n} \int_0^{k_{Fi}(r)}
p^2 \d p \frac{M^*}{\epsilon}
$$
with
$\epsilon=\sqrt{p^2+{M^*}^2}$ and
$$
B_i= \int \d^3r \rho_i, \quad
\rho_i(r)= \frac{\gamma}{6\pi^2}k_{Fi}^3(r).$$
\noindent From the above expressions we get for (\ref{Omega})
$$
\Omega= \int \d^3r \left( \frac{1}{2} \left [
(\nabla \phi)^2 -(\nabla V_0)^2 - (\nabla b_0)^2 - (\nabla A_0)^2
\right] + V_{ef}\right)
$$
with
$$ V_{ef}= \frac{1}{2} \left[
m_s^2 \phi^2 -m_v^2 V_0^2 -m_\rho^2 b_0^2  \right] - \mu_p \rho_p-\mu_n \rho_n
$$
\begin{equation}
+ \frac{\gamma}{2 \pi^2} \sum_{i=p,n} \int_0^{k_{Fi}} \d p p^2 \epsilon +
\Gamma_v V_0 \rho + \Gamma_{\rho} \frac{b_0}{2} \rho_3
+ e A_0 \rho_p 
\label{vef}
\end{equation}

Minimization of $\Omega$ with respect to $ k_{Fi}(r),\, i=p,n$,
gives rise to the following conditions
$$
k_{Fp}^2\left(\mu_p-\sqrt{k_{Fp}^2+{M^*}^2}-
\Gamma_v V_0- \frac{\Gamma_\rho}{2} b_0 -e A_0 - \Sigma_0^R  \right)=0
$$
and
$$
k_{Fn}^2\left(\mu_n-\sqrt{k_{Fn}^2+{M^*}^2}-
\Gamma_v V_0+ \frac{\Gamma_\rho}{2} b_0 - \Sigma_0^R \right)=0,
$$
where the rearrangement term is
$$
\Sigma_0^R=\frac{\partial\, \Gamma_v}{\partial \rho}\, \rho\, V_0+\frac{\partial\, \Gamma_\rho}{\partial \rho}\, \rho_3\, \frac{b_0}{2}-
\frac{\partial\, \Gamma_s}{\partial \rho}\, \rho_s\, \phi .
$$
\noindent From the above equations we obtain $k_{Fp}=0$ and $k_{Fn}=0$ or,
for $k_{Fp}$ or $k_{Fn}$  different from zero,
\begin{equation}
\mu_p=\sqrt{k_{Fp}^2+{M^*}^2}+ \Gamma_v V_0  + \frac{\Gamma_\rho}{2}  b_0 +
e A_0  + \Sigma_0^R, \label{mup}\end{equation}
\begin{equation}
\mu_n=\sqrt{k_{Fn}^2+{M^*}^2}+ \Gamma_v V_0  - \frac{\Gamma_\rho}{2}  b_0
+\Sigma_0^R.
\label{mun}\end{equation}
The values of $k_{Fp}$ and $k_{Fn}$ are obtained inverting these
 two last equations.

Such density dependences in the coupling parameters do not affect
the energy functional but of course affect its derivative such as the
pressure density and the chemical potentials.
As already discussed in the literature \cite{inst04,inst06,inst062,ddpeos},
the rearrangement term is crucial in obtaining different behaviors in
physical properties related to
the chemical potentials or to their derivatives with respect to the
density, such as spinodal regions,  as
compared with the more common NL3 or TM1 parametrizations.

\section{NL${\omega\rho}$ model}

The  Lagrangian density that incorporates the extra non-linear
$\sigma-\rho$ and $\omega-\rho$ couplings
\cite{hp01,hp2001,bunta,bunta2} reads

$$
{\cal L}=\bar \psi\left[\gamma_\mu\left(i\partial^{\mu}-g_v V^{\mu}-
\frac{g_{\rho}}{2}  {\boldsymbol {\tau}} \cdot \mathbf {b}^\mu \right. \right.
$$
$$ \left. \left. -e  \frac{(1+\tau_{i3})}{2} A^\mu \right)
-(M-g_s \phi)\right]\psi
$$
$$
+\frac{1}{2}(\partial_{\mu}\phi\partial^{\mu}\phi
-m_s^2 \phi^2)
-\frac{1}{3!}\kappa \phi ^{3}
-\frac{1}{4!}\lambda \phi ^{4}
-\frac{1}{4}\Omega_{\mu\nu}\Omega^{\mu\nu}$$
$$+\frac{1}{2}m_v^2 V_{\mu}V^{\mu}
-\frac{1}{4}\mathbf B_{\mu\nu}\cdot\mathbf B^{\mu\nu}+\frac{1}{2}
m_\rho^2 \mathbf b_{\mu}\cdot \mathbf b^{\mu}
-\frac{1}{4}F_{\mu\nu}F^{\mu\nu}
$$
\begin{equation}
+g_\rho^2\mathbf b_{\mu}\cdot \mathbf b^{\mu}[\Lambda_s
g_s^2\phi^2+\Lambda_v g_v^2 V_\mu V^\mu],
\label{lagacoplada}
\end{equation}
where $\Omega_{\mu\nu}$, $\mathbf B_{\mu\nu}$ and $F_{\mu\nu}$ are defined after
eq.(\ref{lagtw}).
The  parameters of the model are again the masses and the couplings, which are
now constants, i.e., $g_s$ replaces $\Gamma_s$, $g_v$ replaces $\Gamma_v$ and
$g_\rho$ replaces $\Gamma_\rho$. Non-linear $\sigma$ terms are also included.
We have followed the prescription of \cite{hp01}, where the starting point was
the NL3 parametrization and the $g_\rho$ coupling was adjusted for each value
of the coupling $\Lambda_i$ studied in such a way that for
$k_F=1.15$ fm$^{-1}$ (not the saturation point) the symmetry energy is
25.68 MeV. In the present work we set $\Lambda_s=0$ as in \cite{bunta2}.
Notice that other possibilities for this model with
$\sigma-\rho$ and $\omega-\rho$ couplings have already been discussed in the
literature as in \cite{piek06}, for instance.

The mesonic equations of motion in the Thomas-Fermi approximation become

\begin{equation}
\nabla^2 \phi = m_s^2\phi- g_s \rho_s + \frac{\kappa}{2} \phi^2 +
\frac{\lambda}{6} \phi^3
\label{elphi2}
\end{equation}
\begin{equation}
\nabla^2 V_0 = m_v^2 V_0 - g_v \rho + 2 \Lambda_v g_v^2\, V_0\, g_{\rho}^2
b_0^2,
\label{elV02}
\end{equation}
\begin{equation}
\nabla^2 b_0 =m_\rho^2 b_0 -\frac{g_\rho}{2} \rho_3 + 2 \Lambda_v
g_{\rho}^2 b_0 g_v^2 V_0^2,
\label{elb02}
\end{equation}
\begin{equation}
\nabla^2 A_0 =-e \rho_p,\label{elA02}
\end{equation}

and the expression for the energy reads

$$E= \sum_i \int \d^3r \left(
\gamma \int_0^{k_{Fi}(r)}  \frac{\d^3p}{(2\pi)^3}\,
\sqrt{{\mathbf p}^2 + {M^*}^2} \right.$$
$$+\frac{1}{2} \left[ (\nabla \phi)^2 + m_s^2 \phi^2 
-(\nabla V_0)^2 - m_v^2 V_0^2 \right.$$
$$ - (\nabla b_0)^2 - m_\rho^2 b_0^2  - (\nabla A_0)^2 $$
$$\left. + g_v V_0 \rho + \frac{g_\rho}{2} \rho_3 b_0 + e A_0 \rho_p \right]$$
\begin{equation}
\left. +\frac{\kappa}{6} \phi^3 + \frac{\lambda}{24} \phi^4 
-\Lambda_v g_v^2 V_0^2 g_\rho^2 b_0^2 \right).
\end{equation}

All other expressions are very similar to the ones obtained from the TW model
and can be read off from them bearing in mind that the density dependent
couplings have to be replaced by the constant couplings.  In
  particular the chemical potentials do not contain the rearrangement
  term $\Sigma_0^R$.

\section{Numerical result via a nucleation process}

At this point, eqs. (\ref{elphi}-\ref{elA0}) for the TW model and
eqs. (\ref{elphi2}-\ref{elA02}) for the NL$\omega\rho$ model
have to be solved numerically in a self-consistent way and
hence, initial and boundary conditions for each equation are necessary.
One of the methods we use here is based on a prescription given in
\cite{gotas}, where these conditions are obtained from a situation of phase
coexistence in a mean field approximation with classical meson fields and no
electromagnetic interaction.
The method is well explained in \cite{gotas} and, as we are using different
models here, just the main equations are written next.

For the TW model, the equilibrium equations for homogeneous matter for the
fields are:
\begin{eqnarray}
m_{s}^2\phi
- \Gamma_{s} ~\rho_{s } &=&0, \label{phitw} \\
m_{v}^2 V_0 - \Gamma_{v}~ \rho&=&0, \label{V0tw}\\
m_\rho^2 b_0 - \frac{\Gamma_{\rho}}{2}~ \rho_3&=&0,
\label{b0tw}
\end{eqnarray}
and for the energy and pressure density:
$$
{\cal E}=\frac{1}{\pi^2} \sum_i \int_0^{k_{Fi}} p^2 dp~
\sqrt{{\mathbf p}^2 + {M^*}^2}
$$
\begin{equation}
+\frac{m_{s}^2}{2} \phi^2 + \frac{m_{v}^2}{2} V_0^2
+ \frac{m_{\rho}^2}{2} b_0^2,
\label{enermfa}
\end{equation}
$$
P=\frac{1}{3 \pi^2} \sum_i \int_0^{k_{Fi}} \frac{p^4 dp}{\epsilon}
- \frac{m_{s}^2}{2} \phi^2 + \frac{m_{v}^2}{2} V_0^2
+ \frac{m_{\rho}^2}{2} b_0^2 $$
\beq
+ \rho {\Sigma^R_0}.
\label{pressmfa}
\eeq

For the NL$\omega\rho$ model, the  equilibrium equations  for homogenous matter, energy density and
pressure become:
\begin{equation}
m_s^2\phi- g_s \rho_s + \frac{\kappa}{2} \phi^2 +
\frac{\lambda}{6} \phi^3 =0,
\label{phi2}
\end{equation}
\begin{equation}
m_v^2 V_0 - g_v \rho + 2 \Lambda_v g_v^2\, V_0\, g_{\rho}^2
b_0^2 =0,
\label{V02}
\end{equation}
\begin{equation}
m_\rho^2 b_0 -\frac{g_\rho}{2} \rho_3 + 2 \Lambda_v
g_{\rho}^2 b_0 g_v^2 V_0^2 = 0,
\label{b02}
\end{equation}

$${\cal E}= \frac{1}{\pi^2} \sum_i \int_0^{k_{Fi}}
p^2 dp \sqrt{{\mathbf p}^2 + {M^*}^2}$$
$$+\frac{1}{2} \left[m_s^2 \phi^2 - m_v^2 V_0^2 - m_\rho^2 b_0^2 \right]
+ g_v V_0 \rho + \frac{g_\rho}{2} \rho_3 b_0$$
\begin{equation}
+\frac{\kappa}{6} \phi^3 + \frac{\lambda}{24} \phi^4 
-\Lambda_v g_v^2 V_0^2 g_\rho^2 b_0^2.
\end{equation}
and
$$
P=\frac{1}{3 \pi^2} \sum_i \int_0^{k_{Fi}} \frac{p^4 dp}{\epsilon}
- \frac{m_{s}^2}{2} \phi^2 + \frac{m_{v}^2}{2} V_0^2
+ \frac{m_{\rho}^2}{2} b_0^2 $$
\beq
-\frac{\kappa}{6} \phi^3 -\frac{\lambda}{24} \phi^4 
+ \Lambda_v g_v^2 V_0^2 g_\rho^2 b_0^2.
\eeq

\begin{figure}[b]
\begin{center}
\begin{tabular}{cc}
\includegraphics[width=8.cm]{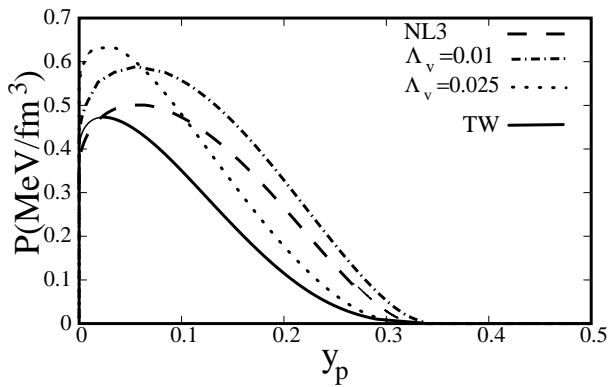} \\
\end{tabular}
\end{center}
\caption{Binodal section for the NL3, TW  and NL$\omega\rho$
parametrizations.}
\label{binodais}
\end{figure}

Based on the geometrical construction and Gibbs conditions for phase
coexistence, i.e., the pressure and both chemical potentials are equal in both
phases, we build the binodal section given in Fig. \ref{binodais}.
Notice that we have defined the proton fraction of the system as
\begin{equation}
y_p = \frac{\rho_p}{\rho}.
\end{equation}
The binodal section yields the boundary conditions which we need. For the same
pressure, two points, with different proton fractions are found. For each of
these points, the meson fields and the densities are well defined and used as
the initial and boundary conditions in eqs. (\ref{elphi}-\ref{elA0}), which
are then solved. Once the meson fields are obtained, all the quantities that
depend on them, as the energy, pressure densities, chemical potentials,
baryonic densities, etc are also computed. The solution is a droplet
with  a certain proton fraction surrounded by a gas of neutrons. If stable
nuclei
are calculated, the gas vanishes because the energy of the system lies below
the neutron drip line and the finite nuclei properties are easily calculated.
This is the general method, but the results depend strongly on the model used
because of the reasons discussed in Section VI.

\section{Numerical result within a harmonic oscillator basis}

Here a different prescription for solving the equations of motion and the
thermodynamical quantities within the Thomas-Fermi approximation is used.
According to \cite{ring}, meson field equations of motion of the Klein-Gordon
type with sources can be carried out by an expansion in a complete set of
basis states. The harmonic oscillator functions with orbital angular momentum
equal to zero are then chosen. The oscillator length is given by
\begin{equation}
b_B=\frac{b_0}{\sqrt 2}, \quad b_0=\sqrt{\frac{\hbar}{M \omega_0}},
\end{equation}
where $M$ is the nucleon mass and $\omega_0$ is the oscillator frequency.
The meson fields and their corresponding inhomogeneous part can be expanded as
\begin{equation}
\Lambda(r)=\sum_{n=1}^{n_B} \Lambda_n R_{n0}(r), \quad
S_\Lambda(r)=\sum_{n=1}^{n_B} S_n^\Lambda R_{n0}(r),
\label{ans}
\end{equation}
where $\Lambda(r)=\phi(r),V_0(r),b_0(r)$ and
\begin{equation}
R_{n l}(r)=\frac{N_{nl}}{b_0^{3/2}} x^l L^{l+1/2}_{n-1}(x^2)exp(-x^2/2),
\end{equation}
where $x=r/b_0$ is the radius measured in units of the oscillator length,
\begin{equation}
N_{nl}=\sqrt{2(n-1)!/(l+n-1/2)!}
\label{norm}
\end{equation}
is the normalization constant and $L^m_n(x^2)$ are the associated Laguerre
polynomials. For the calculation of the meson fields $l=0$ in the expressions
given below.
Once the ansatz given by eqs.(\ref{ans}) are substituted into eqs.(\ref{elphi}-
\ref{elb0}), a set of inhomogeneous equations is obtained:
\begin{equation}
\sum_{n'=1}^{n_B} {\cal H}_{n n'} \Lambda_{n'}=S_n^\Lambda
\end{equation}
where
$$ {\cal H}_{n n'}= \delta_{n n'}\left( b_B^{-2} (2(n-1)+3/2) + m^2_\Lambda
\right)$$
\begin{equation}
+ \delta_{n n'+1} b_B^{-2} \sqrt{n(n+1/2)}+ \delta_{n+1 n'} b_B^{-2}
\sqrt{n'(n'+1/2)}.
\end{equation}
Only the massive fields can be calculated with this method because the 
convergence of the Coulomb field, which has a long range, is very
slow. The Green's function method is then chosen to describe the
electromagnetic interaction:

\begin{equation}
A_{0}(r)=e~\int~r'^{2}dr'\rho_{p}(r')G_c(r,r'), \label{green}
\end{equation}
with
\begin{equation}
G_c(r,r')= \biggl\{ \begin{array}{c}
1/r {~~~~\rm for} ~~r>r'  \\
1/r' {~~~~\rm for} ~r'>r. \end{array}
\end{equation}

\section{Results}

\subsection{Parity Violating Electron Scattering and the Neutron Radius}

We start this section by defining the asymmetry for polarized
electron scattering of a hadronic target as
\begin{equation}
\mathcal{A}=\frac{d\sigma_{+}/d\Omega-d\sigma_{-}/d\Omega}{d\sigma_{+}/d\Omega+d\sigma_{-}/d\Omega},
\end{equation}
where $d\sigma_{\pm}/d\Omega$ is the differential cross section
for initially polarized electrons with positive($+$) and negative
($-$) helicities. As the electromagnetic interaction is not
sensitive to the above difference, the asymmetry becomes dependent
of the weak interaction between the electron and the target.
Moreover, we know from the Standard Model that the neutral Z-boson
couples more strongly to the neutron than to the proton. Those
reasonings were then used in \cite{Don} to first propose a clean
way to determine the neutron distribution in nuclei. If we
consider elastic scattering on an even-even target nucleus,
the asymmetry can be written in the form:

\begin{equation}
\mathcal{A}=\frac{Gq^2}{2\pi\alpha\sqrt{2}}a[\beta_{V}^{p}+\beta_{V}^{n}\frac{\rho_{n}(q)}{\rho_{p}(q)}].\label{ass}
\end{equation}

In the above expression, G, $\alpha$, $a$ and $\beta_{V}^{p,n}$
are Standard Model coupling constants as defined in \cite{Don},
$q$ is the transferred momentum by the electron to the nucleus
and,

\begin{equation}
\rho_{n(p)}(q)=\int~d^{3}r~j_{0}(qr)\rho_{n(p)}(\mathbf {r}),
\end{equation}

\noindent $\rho_{n(p)}(\mathbf {r})$ being the neutron (proton)
distribution in configuration space and $j_{0}$ the spherical
Bessel function of order zero. It is then clear
that a small $q$ measurement of the asymmetry gives the
neutron radius of the distribution once the proton radius is well
known. The proton and neutron mean-square radius are defined as

\begin{equation}
R_{i}^{2}=\frac{\int~d^{3}r r^2 \rho_{i}(\mathbf {r})}{\int~d^{3}r
 \rho_{i}(\mathbf {r})}, \quad i=p,n.
\end{equation}
The neutron skin thickness is defined as
\begin{equation}
\theta=R_n-R_p. \label{skin}
\end{equation}

In the PREX experiment mentioned in the Introduction, the
asymmetry is expected to be measured at $q\approx0.4~fm^{-1}$
\cite{prex}. Also, because the target is a heavy nucleus
($^{208}$Pb), the above results for the asymmetry should be
reconsidered for a detailed comparison with the experiment, since
they were obtained using a Plane Wave Born Approximation for the
electron \cite{horo98}. For our present purposes, eq. (\ref{ass}) is
sufficient to illustrate the sensitivity to the
different model parametrizations and is used next in the
presentation of our numerical results.

The surface energy per unit area of the droplets in the small
surface thickness approximation, excluding the electromagnetic field, reads 
\cite{gotas}
\begin{equation}
\sigma=\int_0^\infty \d r \left[ \left(\frac{\d \phi}{\d r}\right)^2-
\left(\frac{\d V_0}{\d r}\right)^2 -
\left(\frac{\d b_0}{\d r}\right)^2 \right].
\label{sig}
\end{equation}
However,
as the electromagnetic interaction does not contribute to surface properties
directly, we have kept the same definition for the surface energy.

In Table II we show the neutron and proton radius, the neutron skin
  thickness, the binding energy and the surface energy obtained within the
  Thomas-Fermi approximation and the two different numerical prescriptions
  described in the previous sections.
All the results are sensitive to the numerical calculation although the
analytical approximation is the same. When the nucleation method is performed,
the neutron radius is systematically larger, what results in a thicker neutron
skin. This is correlated with the fact that the surface energy is lower within
the nucleation calculation than within the harmonic oscillator method.
Within the same numerical prescription, the neutron skin thickness is smaller
with the TW model than with the NL3. As the coupling strength $\Lambda_v$
increases in the NL$\omega\rho$ model, the results move from the original NL3
to the TW results for all quantities, except the proton radius, which
oscillates a little. We have also included the results obtained with the HS
parametrization \cite{hs} because we have used this parametrization in order
to compare the TF and the Dirac results for the cross sections, as discussed
in the following. As this parametrization is known not to give as good 
  results as the other parametrizations of the NLWM for finite nuclei, we do 
not comment on the results it
provides. Notice that the experimental radius for the protons is
obtained from the charge radius $R_c$ and it is given by
$R_p=\sqrt{R_c^2 -0.64}$ in fm \cite{ring}.
Our results can be compared with
experimental and other theoretical results found in the literature.
The proton radius, which is
known to better than 0.001 fm is better described within the TW model.
This quantity is practically independent of the $\omega-\rho$ interaction
strength in the NL$\omega\rho$ model as far as the HO numerical prescription
is used. The neutron radius, on the other had, is strongly model dependent
with drastic consequences in the neutron skin thickness calculation.
The experimental values for $\theta$ are still very uncertain and 
all our results fall inside the experimental confidence interval.
We shall comment on possible restrictions to the neutron skin thickness in the 
next section. NL3 provides the best results for the binding energy.

In \cite{nl3}, the results shown for the proton and neutron radius are
respectively 5.52 and 5.85 fm, yielding a skin of 0.33 fm, larger than ours.
 Notice, however, that in \cite{nl3} the Dirac equation was explicity solved.
In \cite{piek06}, the authors obtained a value of 0.21 fm for the neutron skin
thickness and a binding energy of -7.89 MeV within a different
parametrization of the NL$\omega\rho$ model.  Again in this case the Dirac
equation was solved.

\begin{figure}[b]
\begin{center}
\begin{tabular}{cc}
\includegraphics[width=8.cm]{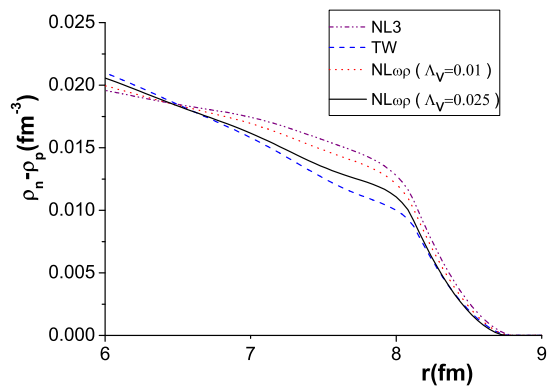} \\
\end{tabular}
\end{center}
\caption{Difference between neutron and proton densities obtained with
the Thomas-Fermi approach solved in a harmonic oscillator basis for the models
discussed in the present work.}
\label{ronp1}
\end{figure}

In Fig. \ref{ronp1} we show the difference between neutron and proton
densities at the $Pb$ surface for the models discussed in the present work 
with the Thomas-Fermi approximation solved in a harmonic oscillator basis.
While the curves deviate a little in between 6.0 and 8.0 fm, at the very
surface they are similar, but a small discrepancy, reflecting the
differences in the neutron skin can be seen.

\begin{figure}[b]
\begin{center}
\begin{tabular}{cc}
\includegraphics[width=8.cm]{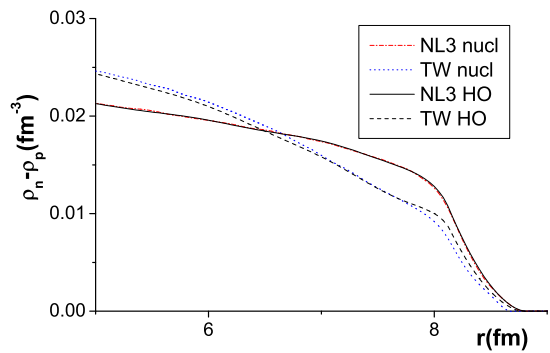} \\
\end{tabular}
\end{center}
\caption{Difference between neutron and proton densities obtained with
the Thomas-Fermi approach solved with both numerical prescriptions for the TW
model.}
\label{ronpcomp1}
\end{figure}

In Fig. \ref{ronpcomp1} we display again the difference between neutron and 
proton densities within both numerical calculations of the TW and NL3 models.
These two Thomas-Fermi calculations should have given more similar results. 
However the nucleation method predicts a very small surface energy for the NL3 
parametrization, and therefore, a large radius. This may be related to the 
choice of the boundary conditions and a deeper comparison between the two 
methods will be pursued.


\begin{figure}[b]
\begin{center}
\begin{tabular}{cc}
\includegraphics[width=8.cm]{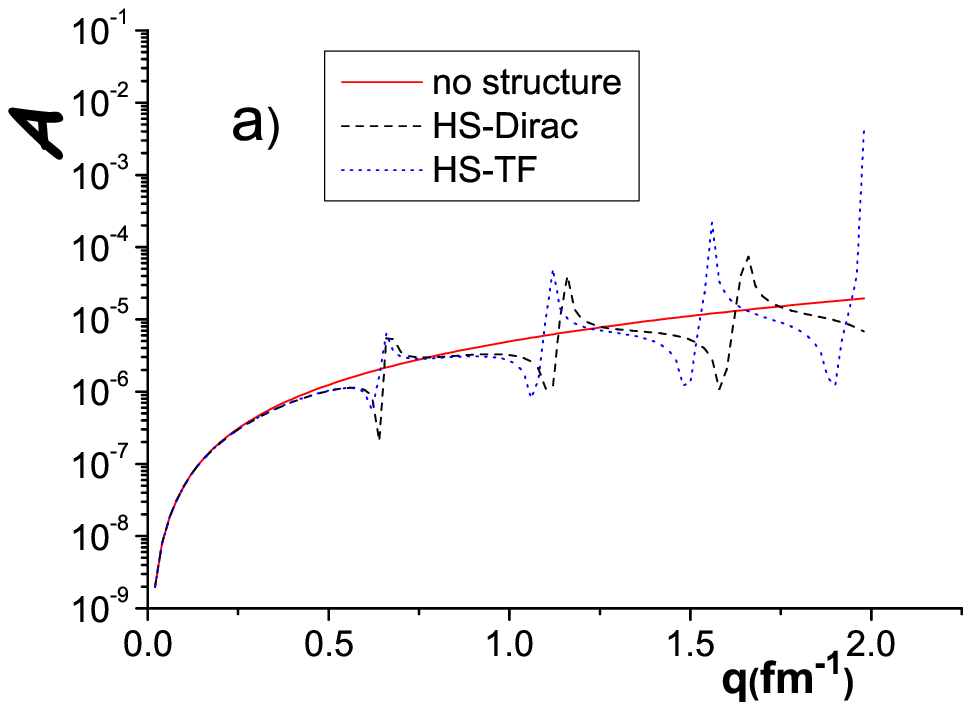} \\
\includegraphics[width=8.cm]{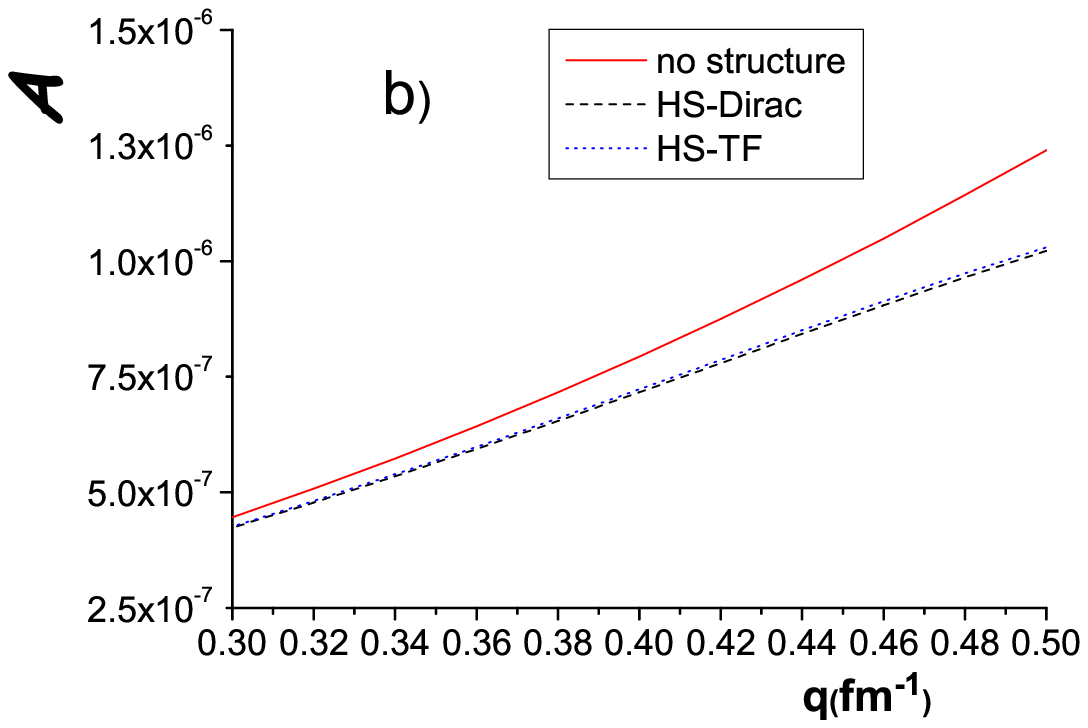} \\
\end{tabular}
\end{center}
\caption{Parametrization HS, comparison Thomas-Fermi-HO versus Dirac-HO}
\label{asym0}
\end{figure}

Next we present our results for the asymmetry given by eq. (\ref{ass}) as a 
function of the transfered momentum. We begin
with Fig.\ref{asym0} which displays the results for the  HS
parametrization of the Walecka model. The curve labeled {\it no
structure} means the case where $Z\rho_{n}(r)=N\rho_{p}(r)$ and
the other two curves are obtained within the TF approximation and
the full solution of the Dirac equation in the Hartree
approximation. At the momentum transfer values of recent
experimental interest (around $\simeq 0.4$ fm$^{-1}$),  the
curves are almost identical. A careful analysis of the same
results in a different scale shows us that the asymmetry changes
$12$ and $11$ percent respectively within the Dirac and TF
approximations in comparison with the {\it no structure} case.
Since it is the measurement of the asymmetry in this low momentum transfer 
region that will provide the accurate result for the neutron skin thickness, 
we have restricted our calculations to the TF approximation, as stated in the
Introduction.

\begin{figure}[b]
\begin{center}
\begin{tabular}{cc}
\includegraphics[width=8.cm]{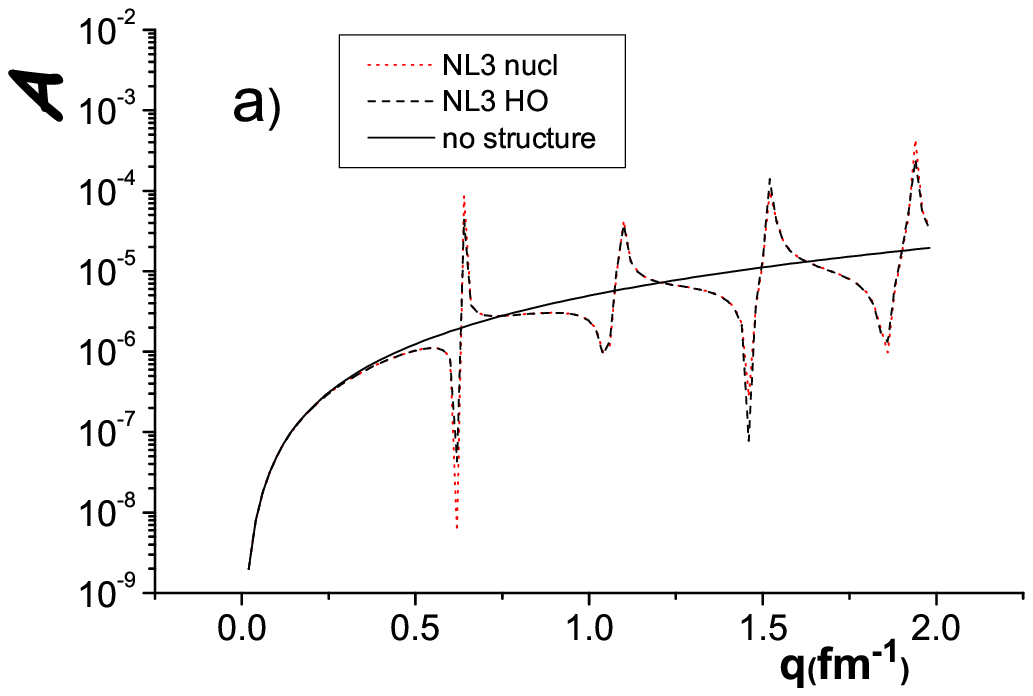}\\
\includegraphics[width=8.cm]{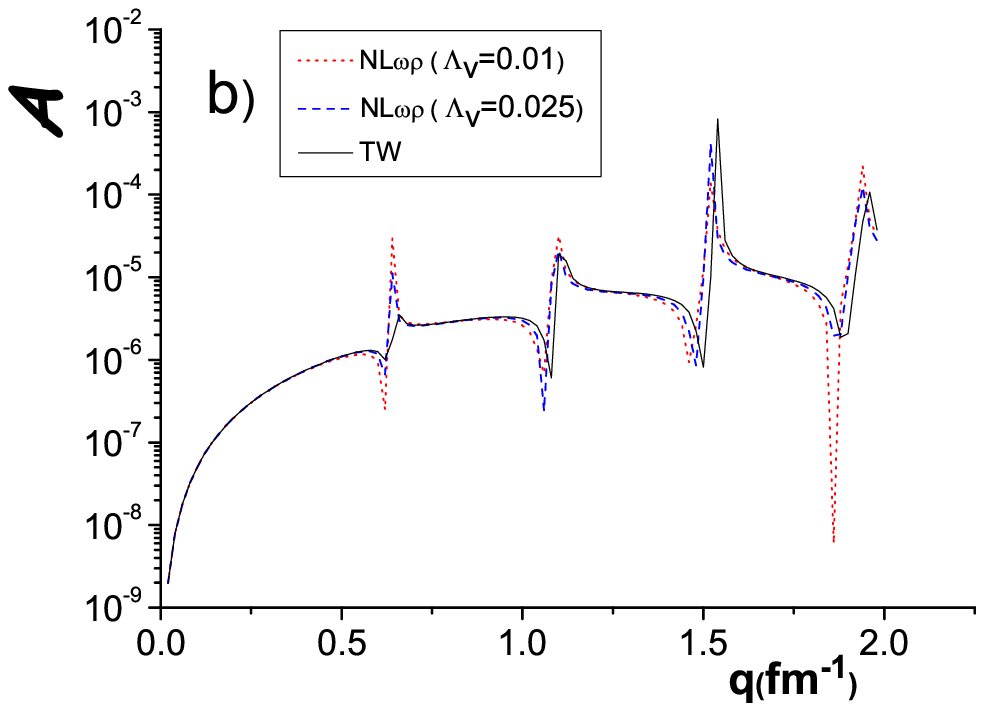} \\
\end{tabular}
\end{center}
\caption{Asymmetry obtained with a) NL3 with both numerical prescriptions and
b)parametrizations NL$\omega\rho$ and TW}
\label{asym2}
\end{figure}

 In Fig. \ref{asym2}a we show the asymmetry obtained with the NL3
model for both numerical calculations in the TF approximation,
i.e, nucleation and HO expansion methods. In this case, the
agreement is very satisfactory even for larger $q$-values, although
the small numerical discrepancies is reflected in a $\sim 10$ percent
difference in the predicted neutron skin thickness, as can be seen
from Table II. Finally, in Fig. \ref{asym2}b our results for the
NL$\omega\rho$ (using two different values for the $\omega-\rho$
coupling constant) and the TW models within the HO numerical
prescription are shown. Again, at low momentum transfers, all
curves coincide. However, it should be noticed that even for two
different model parametrizations which lead us to identical
neutron skin thicknesses, a measurement of the asymmetry in a
higher $q$-region with a modest experimental precision, can
distinguish between them. Also, we should expect that the
asymmetry presents more structure in this high momentum transfer
region if we solve the Dirac equation instead of using the TF
approach, once the high $q$ value region is much more sensitive to
the central part of the neutron distribution, which is known to be
{\it flat} in the TF approximation. These differences can be seen in 
Fig.\ref{asym0}.

\section{Different EoS, different neutron skins}

For the sake of completeness, at this point, we discuss some of the
differences between the TW, the NL$\omega\rho$ models
and the NL3 parametrization of the NLWM.

\noindent From Fig. \ref{binodais} one can see that the largest possible
pressure for a phase coexistence in the TW model is much lower, and appears
at  a lower proton fraction than the NL3 model. This gives rise to a thinner
crust within the TW model, which may imply that the more exotic {\em pasta}
shapes will not form \cite{haen00}. The NL$\omega\rho$ model goes on a
different direction, i.e., the pressure becomes higher than the one obtained
with the NL3 as the $\Lambda_v$ coupling is turned on.

Although the nuclear matter properties fitted to parametrize the models are
quite similar (see Table \ref{bulk}), the way the EoS behaves when
extrapolated to higher or lower densities can vary a lot from a density
dependent hadron model to one of the parametrizations of the NLWM.
Moreover, as seen from Table \ref{bulk}, although the effective mass at
saturation density is lower with the TW than with the NL3,
it can accommodate hyperons if an EoS for stellar matter is necessary, contrary
to the usual NL3 parametrization \cite{ddpeos,alex,compact}.

\begin{table}[h]
\caption{ Nuclear matter properties.}
\label{prop}
\begin{center}
\begin{tabular}{ccccccccccc}
\hline
&  NL3 & & NL$\omega\rho$ & & TW  \\
&   \cite{nl3}  & & \cite{bunta} & & \cite{tw} \\
\hline
& & $\Lambda_v=0.01$ &  $\Lambda_v=0.02$ &  $\Lambda_v=0.025$ & \\
\hline
$B/A$ (MeV) & 16.3 & 16.3 & 16.3 & 16.3 & 16.3 \\
$\rho_0$ (fm$^{-3}$) & 0.148 & 0.148 & 0.148 & 0.148 & 0.153 \\
$K$ (MeV) & 271 & 271   & 271 & 271 & 240 \\
${\cal E}_{sym.}$ (MeV)  & 37.4 & 34.9 & 33.1 & 32.3 & 32.0\\
$M^*/M$ & 0.60 & 0.60 & 0.60 & 0.60 & 0.56 \\
$L$ (MeV)& 118 & 88 & 68 & 61 & 55 \\
$K_{sym}$ (MeV) & 100 & -46 & -53 & -34 & -124\\
\hline
\end{tabular}
\end{center}
\label{bulk}
\end{table}

Another quantity of interest in asymmetric nuclear matter is the nuclear bulk
symmetry energy,
shown in Table \ref{bulk} for the saturation point. The differences in the
symmetry energy at densities larger than the nuclear saturation density is
still not well established, but has already been extensively discussed in the
literature even for the TW model \cite{inst04,inst062,ddpeos,bao-li}.
Again, for the sake of completeness we reproduce these results here because
the neutron skin thickness and the neutron star EoS are related by this
quantity \cite{tb01,steiner,hp01,piek06}, which is usually defined as
$ {\cal E}_{sym} =\left. \frac{1}{2} \frac{\partial^2 {\cal E}/\rho}
{\partial \delta^2} \right|_{\delta=0}$, with
$\delta=-\rho_3/\rho=1-2y_p$. The symmetry energy can be analytically
rewritten as
\begin{equation}
{\cal E}_{sym}= \frac{k_F^2}{6 \epsilon_F}+ \frac{\Gamma_\rho^2}
{8 m_\rho^2} \rho, \label{esym}
\end{equation}
for the TW model and as
\begin{equation}
{\cal E}_{sym}= \frac{k_F^2}{6 \epsilon_F}+ \frac{g_\rho^2}{8 {m^*_\rho}^2} \rho
\label{esymd}
\end{equation}
with the effective $\rho$-meson mass defined as \cite{hp01}
$${m^*_\rho}^2=m_\rho^2+ 2 g_v^2 g_\rho^2 \Lambda_v V_0^2$$
for the NL$\omega\rho$ model. In both cases
\[
k_{Fp}=k_F(1+\delta)^{1/3},\qquad k_{Fn}=k_F(1-\delta)^{1/3},
\]
with $k_F=(1.5 \pi^2\rho)^{1/3}$ and
$\epsilon_F=\sqrt{k_F^2+{M^*}^2}$.  In equations (\ref{esym}) and
(\ref{esymd}) the second term dominates at large densities.  It is
seen that the
non-linear $\rho-\omega$ terms introduce a non-linear density
behavior in the symmetry energy of the NLWM parametrizations such as NL3
and TM1. In TW the non-linear density behavior enters through the
density dependent coupling parameters. These non-linear density
behavior is important because the linear behavior of NL3 and TM1
parametrizations predicts too high
symmetry energy at densities of importance for neutron star matter
which has direct influence on the proton fraction dependence with density.
\begin{figure}[b]
\begin{center}
\begin{tabular}{cc}
\includegraphics[width=8.cm]{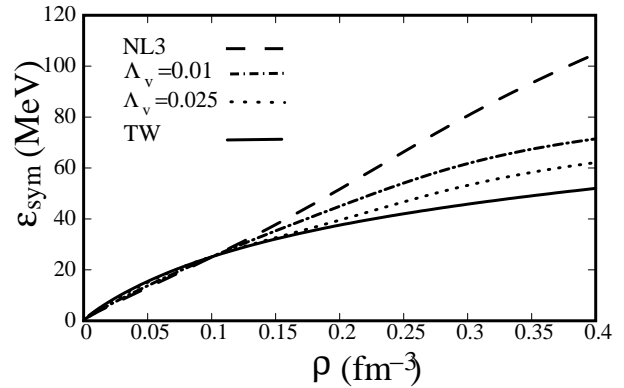} \\
\end{tabular}
\end{center}
\caption{Symmetry energy for the NL3, TW and NL$\omega\rho$ models.}
\label{esymfig}
\end{figure}
\noindent From Fig. \ref{esymfig}, it is easily seen that the symmetry energy
obtained with the TW model behaves in a very different way, as compared with
NL3. In \cite{piek06} a relation between the symmetry energy and the
nuclear binding energy is discussed : the harder the EoS,  the more the
symmetry energy rises with density.
The density dependence discussed in
\cite{piek06} is of the type introduced in \cite{hp01,bunta} through the
inclusion of a $\sigma-\rho$ and/or $\omega-\rho$ couplings and then,
similar with the NL$\omega\rho$ model discussed here. One can observe that
as the strength of the coupling increases, the symmetry energy gets closer to
the TW curve. In fact, in \cite{inst062} it was shown that once this
kind of coupling is introduced with a reasonable strength,
the symmetry energy at low densities tends to behave as the TW model.

\begin{figure}[b]
\begin{center}
\begin{tabular}{cc}
\includegraphics[width=8.cm]{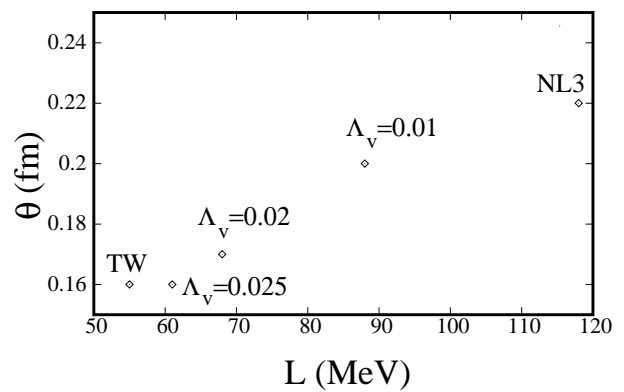} \\
\end{tabular}
\end{center}
\caption{Correlation between the neutron skin $\theta$ and the slope of the 
symmetry energy $L$.}
\label{correlation}
\end{figure}

The symmetry energy can be expanded around the nuclear saturation density
  and reads
\begin{equation}
{\cal E}_{sym}(\rho)={\cal E}_{sym}(\rho_0) +
\frac{L}{3} \left(\frac{\rho-\rho_0}{\rho_0}\right) +
\frac{K_{sym}}{18} {\left(\frac{\rho-\rho_0}{\rho_0}\right)}^2,
\end{equation}
where $L$ and $K_{sym}$ are respectively the slope and the curvature of the
nuclear symmetry energy at $\rho_0$ and they are calculated from
\begin{equation}
L=3 \rho_0 \frac{\partial {\cal E}_{sym}(\rho)}{\partial \rho}|_{\rho=\rho_0}
\quad
K_{sym}=9 \rho_0^2 \frac{\partial^2 {\cal E}_{sym}(\rho)}{\partial^2 \rho}|_
{\rho=\rho_0}.
\end{equation}
These two quantities can provide important information on the symmetry energy
at both high and low densities because they characterize the density
dependence of the energy symmetry. In a recent work \cite{bao-an}, the authors
found a correlation between the slope of the symmetry energy and the neutron
skin thickness. In their work 21 sets of the non-relativistic Skyrme potential
were investigated and only 4 of them were shown to have $L$ values consistent
with the values extracted from experimental isospin diffusion data from heavy
ion collisions. In fact, the extracted value was $L=88 \pm 25$ MeV
\cite{tsang}, which gives a very strong constraint on the density dependence
of the nuclear symmetry energy and consequently on the EoS as well.
A detailed analysis of Table I shows that, if this constraint is to be taken
seriously, neither the NL3 nor the TW model satisfy it. Nevertheless, the
NL$\omega\rho$ slope interpolates beautifully between the NL3 and TW slope
values. Once again it is seen that the increase in $\Lambda_v$ approximates
the NL3 model values for the slope and energy symmetry to the TW values.
Moreover, we have also tried to find a
correlation between the $\theta$ values shown in Table II and $L$ values
displayed in Table I. We found that, as far as some numerical imprecision are
considered, larger values of $L$ correspond to larger values of the neutron 
skin, as seen in Fig. \ref{correlation}.

\begin{figure}[b]
\begin{center}
\begin{tabular}{cc}
\includegraphics[width=8.cm]{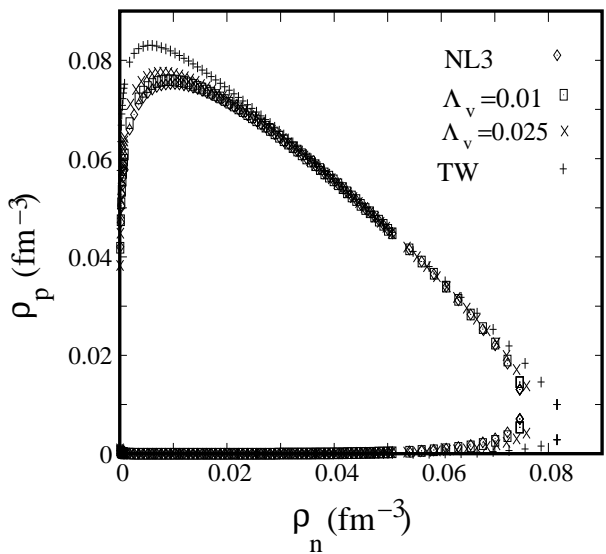} \\
\end{tabular}
\end{center}
\caption{Spinodal section in terms of $\rho_p$ versus $\rho_n$ for the NL3, TW
and NL$\omega\rho$ models.}
\label{spinodais}
\end{figure}

Let's now go back to the problem of solving the differential equations
within the nucleation numerical prescription.
As we need boundary conditions arising from the liquid-gas phase
coexistence in order to solve eqs. (\ref{elphi}-\ref{elA0}) for the TW model
and eqs. (\ref{elphi2}-\ref{elA02}) for the NL$\omega\rho$ model,
the binodal sections are essential and 
the spinodal sections, which separate the regions of stable to
unstable matter are also of interest.
If we had displayed the binodals in a $\rho_p$ versus $\rho_n$ plot, as it is
done with the spinodals in Fig \ref{spinodais}, we could see that the
spinodals surfaces lie inside the binodal sections and share the critical
point corresponding to the highest pressure.

In Fig. \ref{spinodais} the spinodals for
the three different models discussed in this work are shown. Once again, some
of these results can also be found in the recent literature
\cite{inst04,inst062}, but we include them here to make a direct link with the
binodals. The instability of the ANM system  is essentially determined by
density fluctuations in the isoscalar channel. Although the spinodals
  are, by themselves, not relevant in calculations performed at the
  thermodynamical equilibrium, the isospin channel is very sensitive to the
  instabilities occurring below the nuclear saturation density.
The spinodal is determined by
the values of pressure, proton fraction and density for which the determinant
of
\begin{equation}
{\cal F}_{ij}=\left(\frac{\partial^2{\cal F}}{\partial \rho_i\partial\rho_j}
\right)_T,
\label{stability}
\end{equation}
where $\cal F$ is the free energy density, goes to zero. A detailed analysis
of this quantity can be found in \cite{mc03,inst062}.
\noindent {From Fig. \ref{spinodais}, it is seen that the instability
region in the $\rho_p/\rho_n$ plane, defined by the inner section of the
spinodal curve is larger for the TW than for the NL3 model. The size
  of the instability region depends on the derivative of the
  chemical potentials with respect to the neutron and proton densities. At low
densities different models exhibit different behaviors.}

The presence of the rearrangement
term in the TW model also plays a decisive role.
Even though a relatively large  compensation exists between scalar and vector
mesons in the isoscalar channels  within
the rearrangement term at low densities, the spinodal region is
defined by the derivative of the chemical potential and therefore
of the rearrangement term.

Next we examine the spinodals obtained with
different coupling strengths for the NL$\omega\rho$ model. As seen in Fig.
\ref{spinodais}, there is almost no difference between the different
curves. They all fall around the original NL3 curve but once again, they tend
to the TW curve as the coupling strength increases. However,
  contrary to the TW model, it  was shown in \cite{inst06} that the
direction of the instability in NL$_{\omega\rho}$ increases distillation as the
  density increases, and the larger the coupling $\Lambda_v$ the larger
the effect.

 Finally, to end this section, let's make our points clear: we have used a
simple mean field theory approach to obtain the boundary conditions for the
equations of motion of the meson fields in the nucleation prescription.
These boundary conditions depend on
the model used and are intrinsically related with the liquid-gas phase
transition which, in turn, can be well understood by studying
 the coexistence surfaces of the corresponding models.
On the other hand, the neutron skin thickness shows a linear correlation
with the slope of the symmetry energy, as already pointed out in \cite{bao-an}
for non-relativistic models.
 Based on the different behaviors found with density dependent hadronic models
and the NLWM, an obvious consequence is the fact that the neutron skin
thickness depends on the choice of the model.

\section{Conclusions}

We have calculated the $^{208}Pb$ neutron skin thickness with two different
  density dependent hadronic models, the TW and the NL$\omega\rho$ model,
 and one of the most used parametrizations of the NLWM, the NL3. The
  calculations were done within the Thomas-Fermi approximation, which
  gives quite accurate results for the asymmetry in the momentum transfer range
  of interest for the calculation of neutron skins.
In implementing the numerical results two different prescriptions were used:
the first one based on the nucleation process and the second one based on the
  harmonic oscillator basis method. We have seen that when the nucleation 
method is performed,
the neutron radius is systematically larger, what results in a thicker neutron
skin. This is a consequence of the fact that the surface energy is lower within
the nucleation calculation than within the harmonic oscillator method.
Within the same numerical prescription, the neutron skin thickness is smaller
with the TW model than with the NL3. As the coupling strength $\Lambda_v$
increases in the NL$\omega\rho$ model, the neutron skin thickness moves from 
the original NL3 towards the TW results.  We have also found that although the
  neutron skin thickness is  model dependent, the asymmetry at low
  momentum transfers (below 0.5 fm$^{-1}$) is very similar for all models and
  all numerical prescriptions. As $q$ increases, the asymmetry also becomes
  model dependent. The density profiles obtained from the 
solution of the Dirac equation exhibits oscillations near the center of the 
nucleus, behavior which is  not reproduced within the Thomas-Fermi 
approximation.  This fact shows up in the asymmetry at large momentum 
transfers and therefore all the calculations should be
  reproduced by solving the Dirac equation. This calculation is already under
  investigation.

It is worth mentioning that the neutron skin thickness has shown to give hints
on the equations of state that are suitable to describe neutron
stars. Moreover, in \cite{bao-an} a correlation between the slope of the 
symmetry energy and the neutron skin thickness was found for Skyrme-type
models. We have observed that this correlation was also present in the density
dependent models we have studied in the present work. 
 
\section*{ACKNOWLEDGMENTS}

This work was partially supported by CNPq(Brazil),
CAPES(Brazil)/GRICES (Portugal) under project 100/03 and FEDER/FCT (Portugal)
under the projects POCTI/FP/63419/2005 and POCTI/FP/63918/2005.

\newpage

\begin{table}[h]
\caption{$^{208}$ Pb properties}
\label{lead}
\begin{center}
\begin{tabular}{ccccccccccc}
\hline
model & approximation & $R_n$ & $R_p$ & $\theta$ & $B/A$ & $\sigma$ \\
&& (fm) & (fm) & (fm) & MeV & Mev/fm$^2$ \\
\hline
NL3 & TF+nucleation & 5.88 & 5.65 & 0.24 & -7.77 & 0.76 \\
NL3 & TF+HO & 5.79 & 5.57 & 0.22 & -7.79 & 0.96\\
\hline
NL$\omega\rho$, $\Lambda_v=0.01$ & TF+HO & 5.77 & 5.57 & 0.20 & -7.73 & 0.98 \\
NL$\omega\rho$, $\Lambda_v=0.02$ & TF+HO & 5.75 & 5.57 & 0.17 & -7.65 & 0.99
\\
NL$\omega\rho$, $\Lambda_v=0.025$ & TF+HO & 5.74 & 5.58 & 0.16 & -7.63 & 1.00 \\
\hline
TW & TF+nucleation & 5.71 & 5.50 & 0.22 & -6.42 & 1.08\\
TW & TF+HO & 5.68 & 5.52 & 0.16 & -7.46 & 1.10\\
\hline
HS & TF+HO & 5.70 & 5.47 & 0.24 & -6.10 & 1.37\\
\hline
exp.\cite{fricke} & & &  5.44 &  & & \\
exp. \cite{audi}  & & &       &                & -7.87 & \\
exp. \cite{kraszna}& & &      & $0.12\pm 0.07$  &  & \\
exp. \cite{hintz} & & &       & $0.20\pm 0.04$  &  & \\
\hline
\end{tabular}
\end{center}
\end{table}

\end{document}